\documentstyle[prl,aps,psfig,multicol]{revtex}
\newcommand{\be}{\begin{equation}}
\newcommand{\ee}{\end{equation}}
\begin{document}
\title{An Exactly Solvable Anisotropic Directed Percolation Model in Three
Dimensions}
\author{R. Rajesh and Deepak Dhar}
\address{Department of Theoretical Physics, Tata Institute of Fundamental
Research,
Homi Bhabha Road, Bombay 400005, India}
\maketitle
\widetext
\begin{abstract}
We solve exactly a special case of the anisotropic directed bond
percolation problem in three-dimensions, in which the occupation
probability is 1 along two spatial directions, by mapping it to a
$5$-vertex model. We determine the asymptotic shape of the infinite
cluster and hence the direction dependent critical probability.  The
exponents characterizing the fluctuations of the boundary of the wetted
cluster in $d$-dimensions are related to those of the $(d-2)$-dimensional
KPZ equation.
\end{abstract}


\begin{multicols}{2}
It is well-known that the critical behaviour of a very large class of
reaction-diffusion systems is describable in terms of the critical
exponents of directed percolation (DP) \cite{grassberger}. Examples
include heterogeneous catalysis \cite{ziff}, contact processes and
epidemic models, self organized criticality \cite{maslov} and the driven
depinning transition \cite{leschhorn}. Though rather precise estimates of
critical parameters are known from various numerical techniques such as
series expansions and Monte Carlo simulations \cite{jensen}, an exact
solution has not been possible so far, even in the simple case of 1+1
dimensions.  The only analytically tractable case known is the anisotropic
bond percolation on a square lattice, solved by Domany and Kinzel (DK).
They addressed the problem in which horizontal and vertical bonds are
present with different concentrations, and solved exactly the special case
when one of these is set equal to $1$ \cite{domany}. In this case, the
wetted cluster has no holes, and this fact helps in reducing the problem
to that of a simple random walk in one dimension \cite{wu}.

In this paper, we solve the three dimensional generalization of DK model.  
We consider the DP on a three dimensional simple cubic lattice with bond
concentration in the three directions being $p_{x}$, $p_{y}$ and $p_{z}$,
and solve the case $p_{x}=p_{y}=1$, $p_{z}$ arbitrary. We study the
properties of the infinite cluster and obtain the macroscopic shape of the
cluster and the scaling form for the fluctuations of its boundary .  By
universality, the behaviour of fluctuations of the outer boundary of the
infinite cluster at large length scales in this special soluble case is
expected to be the same as in the more general case above percolation
threshold with $p_x, p_y$, and $p_z$ arbitrary.

In addition to the usual critical exponents of the DP problem, which are
defined in terms of the power-law behaviour of various quantities {\it
near} the critical point, there are universal power-law prefactors to the
exponential decay of correlation functions away from the critical point.  
For example, for $p<p_c$, the probability of a finite cluster having $s$
sites varies as $s^{-\theta}exp(-As)$ as $s \rightarrow \infty$. For
$p>p_c$, the probability varies as $s^{-\theta^{\prime}}
exp(-Bs^{\frac{d-1}{d}})$ for large $s$. The exponents $\theta$ and
$\theta^{\prime}$ are examples of {\it off-critical} exponents \cite
{luben}.  We find that the fluctuations of the boundary of the
infinite cluster in d-dimensions are in the same universality class as
those of a (d-2)-dimensional interface moving in a (d-1)-dimensional
space. The latter are described by the well-known KPZ equation \cite
{kardar}. Thus we identify the critical exponents of KPZ equation as
belonging to the class of off-critical exponents of DP.

Our solution falls in the general class of disorder solutions of
statistical mechanical models \cite{rujan}. These often show the phenomena
of dimensional reduction. Thus a 3-dimensional problem at disorder point
can be thought of as a 2-dimensional system evolving in time, and
correlation functions in its steady state would show large distance tails
characteristic of 2-dimensional systems. In our case, the choice
$p_x=p_y=1$ allows a further reduction of dimension by 1, and we show that
the system is equivalent to describing the Markovian evolution of a system
of hard-core particles on a 1-dimensional ring, where the particles are
able to move in only one direction. \cite{footnote2}. Our solution is of
interest also as an exact solution of a nontrivial three-dimensional
statistical mechanics problem with positive weights, of which not many are
known \cite{footnote3}.

Consider the general DP problem on a simple cubic lattice with nearest
neighbor bonds, all directed in the direction of increasing coordinates.
The concentration of bonds is $p_{x},p_{y}$ and $p_{z}$ along the $x$-,
$y$- and $z$- axes. We imagine a source of fluid at the origin, which can
spread along the occupied directed bonds. The sites wetted by this fluid
form a cluster which we shall call the wetted cluster for brevity.  One
would like to calculate the probability, $P(\vec{R})$, that a site
$\vec{R}$ belongs to the wetted cluster. When $p_{x},p_{y}$ and $p_{z}$
are small, the wetted cluster is finite. As the probabilities are
increased, an infinite connected path appears.  We shall call the special
direction along which an infinite path first appears as the preferred
direction.  As the bond concentrations are further increased, percolation
occurs in a narrow cone centered around the preferred direction. Following
DK, we look at the angular dependence of $P(\vec{R})$. Its exponential
decay for large $| \vec{R} |$ defines an angle dependent correlation
length, $\xi_{\Omega}$, in the direction $\Omega$.

The correlation length $\xi_{\Omega}$ diverges as $[p_c(\Omega)-p]^
{-\nu}$, where $p_c(\Omega)$ is the $\Omega$ dependent critical
probability. The value of the exponent $\nu$ depends on the direction
$\Omega$. We show below that in most directions $\nu$ takes the same value
$3/2$.  $\nu$ is different if $\Omega$ is the preferred direction (when
$\nu=\nu_{\|}^{3d}$, the DP exponent), or when it lies in the one of
coordinate planes ($xy,yz$ and $zx$ ).  For percolation in these planes,
clearly all perpendicular bonds can be ignored, and the problem reduces to
the known 2-dimensional case. Consider, for simplicity, the isotropic case
($p_x=p_y=p_z$).  Then the easy direction is (111), and along this
direction $\nu_{\|}^{3d}\approx 1.29$. Along $(110),(011)$ and $(101)$
directions $\nu_{\|}^{2d}\approx 1.733$. Other directions in the $xy,yz$
and $zx$ planes have $\nu=2$. The $x$-, $y$- and $z$- axes have $\nu =1$.

\begin{figure}
\begin{center}
\leavevmode
\psfig{figure=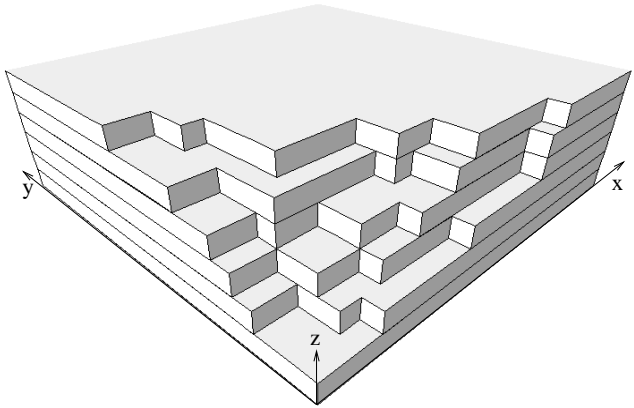,width=8.5cm,angle=0}
\caption{A typical wetted cluster shown up to $z=6$ (for p=0.3). The 
cluster extends infinitely in the $x$ and $y$ directions.}
\label{fig:wet_clust}
\end{center}
\end{figure}

Consider a coordinate system in which $x$- and $y$- axes are in the plane
of the paper with the $z$-axis pointing out of it (see
fig.~\ref{fig:wet_clust}). Since $p_{x}=p_{y}=1$, all the bonds aligned
along the $x$-axis and $y$-axis are present. Hence, if a point $(x,y,z)$
is wetted, then so are all the points $(x^{\prime},y^{\prime},z)$ with
$x^{\prime}\geq x, y^{\prime}\geq y$. It is easy to see that if $(x, y,
z)$ is wet, so must all the points directly below it in the $z$-direction.
Therefore, we can define an integer height function $h(x,y)$ such that all
points $(x,y,z)$ with $z>h(x,y)$ are dry while the points with $z\leq
h(x,y)$ are wet.  The wetted cluster has no holes and can be specified
completely by its bounding surface, $h(x,y)$.

\begin{figure}
\begin{center}
\leavevmode
\psfig{figure=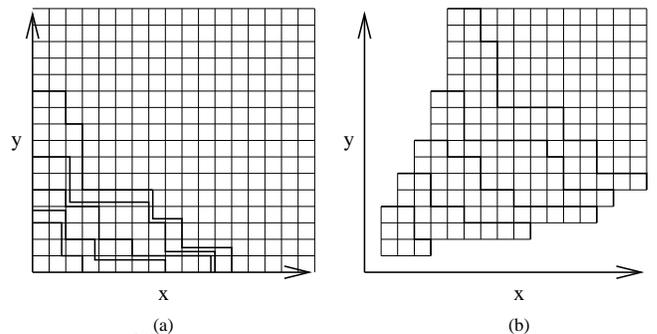,width=8.5cm,angle=0}
\caption{ (a) The projection of the wetted cluster on to the x-y plane.
Overlapping edges have been shown separated for clarity.
(b) The shifted paths.}
\label{fig:projection}
\end{center}
\end{figure}

\begin{figure}
\begin{center}
\leavevmode
\psfig{figure=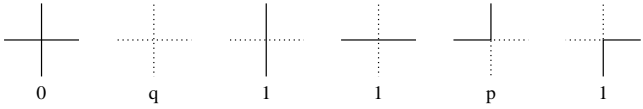,width=8.5cm,angle=0}
\caption{The vertex weights.}
\label{fig:vertex}
\end{center}
\end{figure}

We first determine $\overline{h}(x,y)$, the mean value of $h(x,y)$. We do
so by mapping the problem to a 5-vertex model \cite{footnote4}. Consider
the orthogonal projection of the wetted cluster on to the $xy$ plane (see
fig.~\ref{fig:projection}(a)). It consists of paths running from $y$-axis
to the $x$-axis via steps in the right and down directions obeying the
constraint that paths do not cross each other. The $k^{th}$ path separates
the sites with $h(x,y)<k$ from those sites with $h(x,y)\geq k$.  We now
slide each point in the $k^{th}$ path by $(k,k)$ (see
fig.~\ref{fig:projection}(b)), thereby mapping the point $(x,y)$ to the
point 
$(\zeta,\eta)$ given by
\begin{eqnarray}
\zeta & = & x+h(x,y), \nonumber	\\
\eta & = & y+h(x,y). \label{eq:map}
\end{eqnarray}
This ensures that two paths will not have common edges. Now, every site
has either zero or two bonds connecting it to it's neighbors.  Thus a
given configuration of shifted paths will be made up of $5$ kinds of
vertices. It is easy to check that the correct weights of each of these
vertices are as shown in fig.~\ref{fig:vertex}, where $q=1-p$. Under this
mapping, it is an elementary exercise to show that the gradients are
related through
\begin{eqnarray}
\overline{h}_{\zeta} & = & \overline{h}_{x}\left/ 
\left( 1+ \overline{h}_{x} + \overline{h}_{y}\right)\right. , \nonumber \\
\overline{h}_{\eta} & = & \overline{h}_{y}\left/ 
\left(1+ \overline{h}_{x} + 
\overline{h}_{y}\right)\right. , \label{eq:calculus}
\end{eqnarray}
where $\overline{h}_{\zeta}$ stands for $\frac{\partial
\overline{h}}{\partial \zeta}$, keeping the second coordinate $\eta$
fixed, and similarly for other partial derivatives.

Consider a point $(\zeta,\eta)$ with $\zeta,\eta \gg 1$. In the 5-vertex
problem, $\overline{h}_{\zeta}$ and $\overline{h}_{\eta}$ are the mean
densities of lines in the horizontal and vertical directions respectively.
They are, however, not independent and knowledge of one determines the
other. The relation between $\overline{h}_{\zeta}$ and
$\overline{h}_{\eta}$ is a local relation and is most easily determined
using a different set of boundary conditions when $\overline{h}_{\zeta}$
and $\overline{h}_{\eta}$ are uniform everywhere. Thus we consider the
5-vertex model on an open cylinder of length $N$ in the $y$ direction, and
infinite in the $x$ direction.

\begin{figure}
\begin{center}
\leavevmode
\psfig{figure=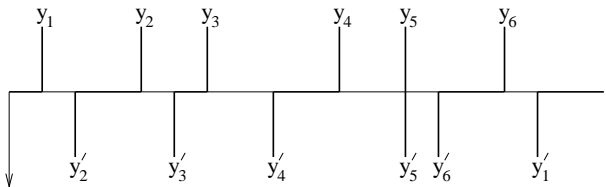,width=8cm,angle=0}
\caption{Two typical arrangements of lines in adjacent column (for $n=6$). 
The position of lines, $\{y_i\}$, in the upper column must interlace with 
the position, $\{y^{\prime}_i\}$, in the lower column.}
\label{fig:transfer}
\end{center}
\end{figure}

The 5-vertex model can be solved via the transfer matrix technique. We
transfer from column to column along the $x$-axis. Consider the sector in
which there are $n$ lines in a column. The ice-rule plus periodic boundary
conditions ensure that there are $n$ lines in each column \cite{lieb1}.
The transfer matrix, $T(\{y^{\prime}\};\{y\})$, which is the weight of
going from configuration $\{y\}$ to $\{y^{\prime}\}$ (see
fig.~\ref{fig:transfer} where the labelling is self explanatory) is the
product of the weights of each vertex and can be written as
\be
T(\{y^{\prime}\};\{y\})=\prod_{i=1}^{n}(p^{1-\delta_{y^{\prime}_{i},y_{i}}}
q^{y^{\prime}_{i}-y_{i-1}-1}).
\ee

It is easy to verify that $\sum_{\{y^{\prime}\}} T(\{y^{\prime}\};
\{y\})=1$, where the summation is over all allowed configurations.  
Therefore $T(\{y^{\prime}\};\{y\})$ is a properly normalized transition
probability of $(\{y\}\rightarrow\{y^{\prime}\})$.

One can visualize this process as a system of $n$ hard core particles,
each hopping only to the right on a ring of size $N$. If the particle on
the right is at a distance $m$ then the particle can hop up to $m-1$ steps
during one time step, with the probability for $k$ steps being
$P(k|m)=p^{1-\delta_{k0}}q^{m-k-1}$. For large times, the above, being a
Markov process, will evolve into its steady state. The steady state will
be one in which all states are equally likely.  To see this, we note that
$\sum_{\{y\}}T(\{y^{\prime}\};\{y\})=1$.  Therefore $T^{t}$, the transpose
of $T$, is also a stochastic matrix.  Hence it has the state
$(1,\ldots,1)$ as the left eigenvector with eigenvalue $1$.  Taking
transpose, $(1,\ldots,1)^{t}$, is a right eigenvector with eigenvalue $1$.

Knowing the steady state, we can determine the average number of steps,
$\overline{d}$, of a particle in the $y$ direction for each transfer in
the x direction.  Simple algebra gives
\be
\overline{d}_{\rho} = \frac {p(1-\rho)}{\rho (p+q \rho)}, \label{eq:rho_sl}
\ee
where $\rho=n/N$. Making the identification $\overline{d}_{\rho}=
\overline{h}_{\zeta}/ \overline{h}_{\eta}$ and $\rho=\overline{h}_{\eta}$
in eq.(\ref{eq:rho_sl}), we get
\be
p \left( 1-\overline{h}_{\zeta} -\overline{h}_{\eta}\right) = 
q \overline{h}_{\zeta} \overline{h}_{\eta}. \label{eq:intermediate}
\ee
Eq.(\ref{eq:intermediate}) can be rewritten in $x,y$ coordinates with the
help of eq.(\ref{eq:calculus}) to give
\be
p \left( 1+\overline{h}_{x} +\overline{h}_{y}\right) = 
q \overline{h}_{x} \overline{h}_{y}. \label{eq:final}
\ee
For large $x,y$, eq.(\ref{eq:final}) has the scaling solution 
\be
\overline{h}(x,y)=\frac{1}{\Lambda}\overline{h}(\Lambda x,\Lambda y).
\ee
Putting the ansatz $\overline{h}(x,y)=A(x+y)+B\sqrt{xy}$, we find that it
satisfies the equation with $A=p/q$ and $B=2\sqrt{p}/q$. 
Thus the asymptotic shape of the surface is given by
\be
\overline{h}(x,y,p) = \left(px+py+2\sqrt{pxy}\right)/q,~for~x,~y \gg 1. 
\label{eq:section}
\ee
The expression is symmetric in $x$ and $y$. If $y=0$, then
$\overline{h}=px/q$ as expected. For given $p$, and given the azimuthal
angle, $\phi$, there is a critical polar angle $\theta_c$ such that all
points with $\theta>\theta_c$ are wetted. The critical angle is determined
through the relation $\cot (\theta_c)=\overline{h}/\sqrt{x^2+y^2}$.  
Eq.(\ref{eq:section}) can then be inverted to derive the direction
dependent critical probability, the smallest probability for which
$(\theta,\phi)$ is wetted, as
\be
p_{c}(\theta,\phi)=\left(\frac{\sqrt{(\alpha+1)(\beta+1)}-\sqrt{\alpha\beta}}
{\alpha+\beta+1}\right)^{2},
\ee
where $\alpha=\tan (\theta)\cos(\phi)$ and $\beta=\tan (\theta) \sin(\phi)$.

The problem can also be viewed as a 1-dimensional cellular automaton
evolving in time.  Consider the coordinate transformation $t=x+y,u=x-y$.
Then from the definition of the model it is easy to check that for $t\geq
0$, $h(u,t)$ follows the rule,
\be
h(u,t+1)=max[h(u-1,t),h(u+1,t)]+\eta(u,t), 
\ee
where $\eta(u,t)'s$ are i.i.d. random variables taking non negative
integer values with $Prob(\eta=k)=(1-p)p^{k}$. The rules of evolution are
local and we expect the fluctuations to be governed by the KPZ equation
\cite{kardar}

We now look at the correlations in the system.  There are two correlation
lengths of interest. $\xi_{\|}(\Omega)$ has been defined through the
exponentially decaying probability of wetting. $\xi_{\|}(\Omega)$ diverges
as $[p_c(\Omega)-p]^{-\nu}$.  We denote by $\xi_{\bot}$ the correlation
length along equal height contours but along the wetted surface. We can
study the fluctuations by looking at the scaling properties of the
5-vertex model. The transfer matrix can be diagonalised via the Bethe
ansatz \cite{lieb2}. From this analysis it is known that the dynamic
exponent $z=3/2$, and the roughening exponent $\chi=1/2$ \cite{dhar}.
These exponents are the well known exponents of the KPZ equation.

The height fluctuations at a point are linearly related to the
fluctuations in the trajectories in the 5-vertex problem. In the KPZ
problem, it is known that the magnitude of the fluctuations of height
varies as $h^{1/3}$. Thus we get $\sqrt{\left<(\delta h)^{2}\right>} \sim
h^{1/3}$. $\xi_{\|}$ is the correlation length at an angle $\theta_{c} +
\delta \theta$. Clearly $\xi_{\|} \delta \theta \sim h^{1/3} \sim
\xi_{\|}^{1/3}$ or $\xi_{\|} \sim (\delta \theta)^{-3/2}$.  From
eq.(\ref{eq:section}), we note that, for $p\neq 0$, $\delta \theta \sim
\delta p$. Thus we get $\nu = 3/2$, to be compared with the value $\nu=2$
for the two-dimensional DK.  The behaviour of $\xi_{\bot}$ can be obtained
from the relation $\xi_{\bot} \sim \xi_{\|}^{1/z}$. Thus we get that
$\xi_{\bot} \sim x_{\|}^{2/3}$, where $x_{\|}$ is the distance of the
height contour from the origin.

Other 3-dimensional lattices can be treated similarly. We consider
lattices made up of two-dimensional layers stacked on top of each other
with vertical bonds connecting a site to the one vertically above it. If
each layer is a triangular lattice, then it may be described as a square
lattice with one set of diagonal bonds. However, the diagonal bonds
provide no additional connections if $p_x = p_y =1$, and can be neglected.
Thus the problem reduces to the square lattice. If each layer is a
honeycomb lattice, there are two types of sites: those having only $2$ or
$3$ outgoing bonds.  If we integrate over the former, the problem again
reduces to a simple cubic lattice, with a renormalized probability $p_z$.

In higher dimensions, $d>3$, the exponents are related to the higher
dimensional KPZ equation. The $d$-dimensional DK model is equivalent to a
$(d-2)$-dimensional interface growing in time. As in the case of depinning
transition of a driven tilted interface \cite{tang}, the height-height
correlation function in $d$-dimensions will have the scaling form
\be
\left<\delta h(x) \delta h(x^{\prime})\right>= |x_{\|}-x_{\|}^{\prime}|
^{\frac{2\chi_{d-2}}{z_{d-2}}} F\left(\frac{|x_{\bot}-x_{\bot}^{\prime}|^
{z_{d-2}}} {|x_{\|}-x_{\|}^{\prime}|}\right),
\ee
where $x_{\|}$ is measured along the radial direction, $x_{\bot}$ is
measured along the equal height contour and $z_{d}$ and $\chi_{d}$ are KPZ
exponents of a growing $d$-dimensional interface.

In summary, we have been able to exactly solve the three-dimensional
anisotropic directed percolation problem for the special case
$p_{x}=p_{y}=1$. We used the fact that in this case the wetted cluster has
no holes within. This reduces the problem to studying the two-dimensional
surface of the cluster. The weights of different surfaces were shown to be
same as those of different configurations of a $5$ vertex model. From the
already known exact solution of the latter, we obtained an exact
expression for the average height profile $h(x,y)$ and determined the
asymptotic correlations of the surface fluctuations of the wetted cluster
for large separations.

We thank Mustansir Barma and Satya Majumdar for critical reading of the
manuscript.

\end{multicols}{2}
\end{document}